# Acoustic and optical variations during rapid downward motion episodes in the deep north-western Mediterranean Sea

H. van Haren[z,*], I. Taupier-Letage[ah,1], J.A. Aguilar[a], A. Albert[b], M. Anghinolfi[c], G. Anton[d], S. Anvar[e], M. Ardid[f] , A.C. Assis Jesus[g], T. Astraatmadja[g,2], J-J. Aubert[h], R. Auer[d], B. Baret[i], S. Basa[j], M. Bazzotti[k,ℓ], V. Bertin[h], S. Biagi[k,ℓ], C. Bigongiari[a], M. Bou-Cabo[f] , M.C. Bouwhuis[g], A. Brown[h], J. Brunner[h,3], J. Busto[h], F. Camarena[f], A. Capone[m,n], G. Carminati[k,ℓ,4], J. Carr[h], D. Castel[b], E. Castorina[o,p], V. Cavasinni[o,p], S. Cecchini[ℓ,q], Ph. Charvis[r], T. Chiarusi[ℓ], M. Circella[s], R. Coniglione[t], H. Costantini[c], N. Cottini[u], P. Coyle[h], C. Curtil[h], G. De Bonis[m,n], M.P. Decowski[g], I. Dekeyser[v], A. Deschamps[r], C. Distefano[t], C. Donzaud[i,w], D. Dornic[h,a], D. Drouhin[b], T. Eberl[d], U. Emanuele[a], J-P. Ernenwein[h], S. Escoffier[h], F. Fehr[d], V. Flaminio[o,p], K. Fratini[x,c], U. Fritsch[d], J-L. Fuda[v], G. Giacomelli[k,ℓ], J.P. Gómez-González[a], K. Graf[d], G. Guillard[y], G. Halladjian[h], G. Hallewell[h], A.J. Heijboer[g], Y. Hello[r], J.J. Hernández-Rey[a], J. Hößl[d], M. de Jong[g,2], N. Kalantar-Nayestanaki[aa], O. Kalekin[d], A. Kappes[d], U. Katz[d], P. Kooijman[g,ab,ac], C. Kopper[d], A. Kouchner[i], W. Kretschmer[d], R. Lahmann[d], P. Lamare[e], G. Lambard[h], G. Larosa[f] , H. Laschinsky[d], D. Lefèvre[v], G. Lelaizant[h], G. Lim[g,ac], D. Lo Presti[ad], H. Loehner[aa], S. Loucatos[u], F. Lucarelli[m,n], K. Lyons[y], S. Mangano[a], M. Marcelin[j], A. Margiotta[k,ℓ], J.A. Martinez-Mora[f], G. Maurin[u], A. Mazure[j], M. Melissas[h], T. Montaruli[s,ae], M. Morganti[o,p], L. Moscoso[u,i], H. Motz[d], C. Naumann[u], M. Neff[d], R. Ostasch[d], G. Palioselitis[g], G.E. Păvălaş[af], P. Payre[h], J. Petrovic[g], P. Piattelli[t], N. Picot-Clemente[h], C. Picq[u], R. Pillet[r], V. Popa[af], T. Pradier[y], E. Presani[g], C. Racca[b], A. Radu[af], C. Reed[h,g], G. Riccobene[t], C. Richardt[d], M. Rujoiu[af], G.V. Russo[ad], F. Salesa[a], F. Schoeck[d], J-P. Schuller[u], R. Shanidze[d], F. Simeone[n], M. Spurio[k,ℓ], J.J.M. Steijger[g], Th. Stolarczyk[u], C. Tamburini[v], L. Tasca[j], S. Toscano[a], B. Vallage[u], V. Van Elewyck[i], M. Vecchi[m], P. Vernin[u], G. Wijnker[g], E. de Wolf[g,ac], H. Yepes[a], D. Zaborov[ag], J.D. Zornoza[a], J. Zúñiga[a]

[a]IFIC - Instituto de Física Corpuscular, Edificios Investigación de Paterna, CSIC - Universitat de València, Apdo. de Correos 22085, 46071 Valencia, Spain
[b]GRPHE - Institut universitaire de technologie de Colmar, 34 rue du Grillenbreit BP 50568 - 68008 Colmar, France
[c]Dipartimento di Fisica dell'Università e Sezione INFN, Via Dodecaneso 33, 16146 Genova, Italy
[d]Friedrich-Alexander-Universität Erlangen-Nürnberg, Erlangen Centre for Astroparticle Physics, Erwin-Rommel-Str. 1, D-91058 Erlangen, Germany
[e]Direction des Sciences de la Matière - Institut de recherche sur les lois fondamentales de l'Univers Service d'Electronique des Détecteurs et d'Informatique, CEA Saclay, 91191 Gif-sur-Yvette Cedex, France
[f] Institut d'Investigación per a la Gestío Integrada de les Zones Costaneres (IGIC) - Universitat Politècnica de València. C/Paranimf, 1. E-46730 Gandia, Spain
[g]FOM Instituut voor Subatomaire Fysica Nikhef, Science Park 105, 1098 XG Amsterdam, the Netherlands
[h]CPPM - Centre de Physique des Particules de Marseille, CNRS/IN2P3 et Université de la Méditerranée, 163 Avenue de Luminy, Case 902, 13288 Marseille Cedex 9, France
[i]APC - Laboratoire AstroParticule et Cosmologie, UMR 7164 (CNRS, Université Paris 7 Diderot, CEA, Observatoire de Paris) 10, rue Alice Domon et L´eonie Duquet 75205 Paris Cedex 13, France
jLAM - Laboratoire d'Astrophysique de Marseille, CNRS/INSU et Université de Provence, Traverse du Siphon - Les Trois Lucs, BP 8, 13012 Marseille Cedex 12, France
[k]Dipartimento di Fisica dell'Università e Sezione INFN, Viale Berti Pichat 6/2, 40127 Bologna, Italy
[ℓ]INFN-Sezione di Bologna, Viale Berti Pichat 6/2, 40127 Bologna, Italy

[m]Dipartimento di Fisica dell'Università La Sapienza, P.le Aldo Moro 2, 00185 Roma, Italy
[n]INFN-Sezione di Roma, P.le Aldo Moro 2, 00185 Roma, Italy
[o]Dipartimento di Fisica dell'Università, Largo B. Pontecorvo 3, 56127 Pisa, Italy
[p]INFN-Sezione di Pisa, Largo B. Pontecorvo 3, 56127 Pisa, Italy
[q]INAF-IASF, via P. Gobetti 101, 40129 Bologna, Italy
[r]GéoSciences Azur, CNRS/INSU, IRD, Université de Nice Sophia-Antipolis, Université Pierre et Marie Curie - Observatoire Océanologique de Villefranche, BP48, 2 quai de la Darse, 06235 Villefranche-sur-Mer Cedex, France
[s]INFN-Sezione di Bari, Via E. Orabona 4, 70126 Bari, Italy
[t]INFN - Laboratori Nazionali del Sud (LNS), Via S. Sofia 44, 95123 Catania, Italy
[u]Direction des Sciences de la Matière - Institut de recherche sur les lois fondamentales de l'Univers-Service de Physique des Particules, CEA Saclay, 91191 Gif-sur-Yvette Cedex, France
[v]COM - Centre dOcéanologie de Marseille, CNRS/INSU et Université de la Méditerranée, 163 Avenue de Luminy, Case 901, 13288 Marseille Cedex 9, France
[w]Université Paris-Sud 11 - Département de Physique - F - 91403 Orsay Cedex, France
[x]Dipartimento di Fisica dell'Università, Via Dodecaneso, 16146 Genova, Italy
[y]IPHC-Institut Pluridisciplinaire Hubert Curien, Université de Strasbourg et IN2P3/CNRS, 23 rue du Loess, BP 28, 67037 Strasbourg Cedex 2, France
[z]Royal Netherlands Institute for Sea Research (NIOZ), Landsdiep 4, 1797 SZ 't Horntje (Texel), the Netherlands
[aa]Kernfysisch Versneller Instituut (KVI), University of Groningen, Zernikelaan 25, 9747 AA Groningen, the Netherlands
[ab]Universiteit Utrecht, Faculteit Betawetenschappen, Princetonplein 5, 3584 CC Utrecht, the Netherlands
[ac]Universteit van Amsterdam, Institut voor Hoge-Energiefysika, Science Park 105, 1098 XG Amsterdam, the Netherlands
[ad]Dipartimento di Fisica ed Astronomia dell'Università, Viale Andrea Doria 6, 95125 Catania, Italy
[ae]University of Wisconsin – Madison, 53715, WI, USA
[af]Institute for Space Sciences, R-77125 Bucharest, Mágurele, Romania
[ag]ITEP - Institute for Theoretical and Experimental Physics, B. Cheremushkinskaya 25, 117218 Moscow, Russia
[ah]Laboratoire d'Oceanographie Physique et de Biogeochimie (LOPB), CNRS UMR 6535 Université de la Mediterranée, Centre d'Oceanologie de Marseille, Antenne de Toulon c/o IFREMER, BP 330, F-83507 La Seyne, France

[*] Corresponding author. E-mail: hans.van.haren@nioz.nl
[1]Not member of the ANTARES collaboration, but contributor to this paper.
[2]Also at University of Leiden, the Netherlands
[3]On leave at DESY, Platanenallee 6, D-15738 Zeuthen, Germany
[4]Now at at the University of California, Irvine, 92697, CA, USA.

**ABSTRACT**

An Acoustic Doppler Current Profiler (ADCP) was moored at the deep-sea site of the ANTARES neutrino telescope near Toulon, France, thus providing a unique opportunity to compare high-resolution acoustic and optical observations between 70 and 170 m above the sea bed at 2475 m. The ADCP measured downward vertical currents of magnitudes up to 0.03 m $s^{-1}$ in late winter and early spring 2006. In the same period, observations were made of enhanced levels of acoustic reflection, interpreted as suspended particles including zooplankton, by a factor of about 10 and of horizontal currents reaching 0.35 m $s^{-1}$. These observations coincided with high light levels detected by the telescope, interpreted as increased bioluminescence. During winter 2006 deep dense-water formation occurred in the Ligurian subbasin, thus providing a possible explanation for these observations. However, the 10-20 days quasi-periodic episodes of high levels of acoustic reflection, light and large vertical currents continuing into the summer are not direct evidence of this process. It is hypothesized that the main process allowing for suspended material to be moved vertically later in the year is local advection, linked with topographic boundary current instabilities along the rim of the 'Northern Current'.

## 1. Introduction

In the past, a large effort has been put in oceanographic studies on mesoscale phenomena like eddies, meandering boundary currents and their effects on the distribution of marine life with abundance along their rims (fronts). Most of these studies focused on the euphotic zone, say the upper few 100 m below the surface. Great help proved satellite imagery of the sea surface in addition to more classic observations from ships. Partly due to logistic problems, less is known about the deep-sea, below 1000 m. Although it is known that Gulf Stream eddies have large vertical extent >2000 m (Richardson, 1983), no direct observations have been made of vertical motions linking abundant near-surface waters with the deep. Recently, Rivas et al. (2009) presented partially such [rare] observations for Gulf of Mexico eddies. If such link exists, it will locally facilitate transport of plankton and food supply to deep-sea organisms living in the dark, where most fauna use light for communication and predation.

In general, bioluminescent organisms are progressively less abundant at greater depths (Vinogradov, 1961; Bradner, 1987; Priede et al., 2006; Heger et al., 2008). Faunal groups that produce bioluminescence in the deep sea are fish and zooplankton (Haddock et al., 2010). Their light is produced by themselves and rarely by symbioting bacteria. In the Mediterranean Sea their presence is a factor of about 10 less abundant than, e.g., parts of the North-Atlantic Ocean, between 1500 and 2500 m and across the Mediterranean values may differ by more than a factor of 10 as a function of time, location and depth (Priede et al., 2008). Most of previous bioluminescence observations come from vertical profiling instruments at different locations. Time series observations from a single location are rare mainly because of power supply issues in self-contained instrumentation. An opportunity for such time series observations is offered by astrophysicists who recently built the ANTARES detector, one of few deep-sea cabled networks.

The ANTARES site is off the French Provençal coast in the north-western Mediterranean Sea, about 10 km from the nearly flat base of the steep continental slope (Fig. 1). It is at the north-eastern edge of the Provençal subbasin (far north-west of Ligurian subbasin). A 40 km long electro-optical cable provides power and the connection for data transmission to and from a shore station (Aguilar et al., 2007).

The ANTARES detector (Ageron et al., 2010) is designed to search for high-energy neutrinos coming from galactic and extra-galactic astrophysical sources for a period of at least 15 years. It is at great depths, mainly to have the water act as a shield for sunlight and cosmic rays, and also to avoid large levels of bioluminescence. The detection principle is based on the collection of Cherenkov photons induced by relativistic charged particles, produced in neutrino interactions, using a 3D-array of about 900 Photo-Multiplier Tubes (PMTs) sensitive to single photo-electrons (Amram et al., 2002). Each PMT together with electronics is integrated in a pressure resistant glass sphere of 17 inch diameter to form an Optical Module (OM). The OMs are mounted on 12 mooring lines and are positioned between about 1900 and 2400 m vertically and 60-70 m apart horizontally. An extra line is used for seismic and oceanographic observations including those on water motions, marine biology and sedimentology. Thus, for long periods of time processes may be studied at a site just off a continental slope into the abyss of a subbasin.

One of potential processes is the Northern Current (NC; Millot, 1999), which flows counter-clockwise along the boundary slopes of the Ligurian and Provençal subbasins and which is driven by buoyancy forces affected by rotation (Crépon et al., 1982; 1989). It is a few tens of km wide and it shows a marked seasonal variability. During summertime the NC is wider (up to ~50 km) and shallower (150-200 m at the slope). During wintertime the NC is narrower (20-30 km) and deeper (200-400 m), and its mesoscale activity increases, generating meanders mostly with a 10-20 day periodicity (e.g., Albérola et al. 1995) and, generally in January, intense mesoscale activity can reach from surface to bottom. This intensification is due to dense water formation in the subbasin (Crépon et al., 1982; 1989). Then, from the NC-rim and into the center of the subbasins vigorous horizontal motions, with speeds up to 0.5 m $s^{-1}$ compared to $O(10^{-2})$ m $s^{-1}$ in summer (Taupier-Letage and Millot, 1986), may become vertically uniform barotropic (Albérola et al. 1995). These are associated with enhanced near-bottom particle fluxes (Martin et al., 2010). As the rim of the NC is meandering, it skims over the ANTARES site so that the detector lies alternately under NC- or under offshore-influence.

Along the NC-rim near the surface, enhanced levels of near-surface phytoplankton grow in spring and may be transported downward along the front, although evidence was so far limited to the upper few 100 m (e.g., Boucher et al., 1987; Gorsky et al., 2002) and more recently to 1000 m (Stemmann et al., 2008). Molinero et al. (2008) and Craig et al. (2010) reported a correlation between surface phytoplankton concentration and the density of bioluminescent (zooplankton) organisms, which are thus also abundant along the NC-rim. More in general, numerous near-surface studies showed aggregation of zooplankton along the edges of eddies and fronts, e.g., Piontkovski et al. (1995), Hernández-León et al. (2001), Jiang et al. (2007), Labat et al. (2009). Presently unknown is the influence of the NC in transporting downward suspended materials including zooplankton to great depths. So far, no direct observations have been reported of the effects of such vertical currents on deep biomass, but a patch of elevated bioluminescence was observed at about 1000 m underneath a mesoscale eddy in the Atlantic (Heger et al., 2008).

In the western basin of the Mediterranean, there may be two physical processes that are characterized by asymmetric, larger down- than upward, w having O($10^{-2}$ m $s^{-1}$) surface-to-bottom magnitudes (Millot, 1999). In these processes, the larger magnitude downward motions are found in areas of smaller horizontal extent than those of the upward motions. Both can affect the ANTARES site:

i) deep dense water convection due to evaporation and cooling of near-surface waters mixing with intermediate waters below, which is predominantly known to occur off the shelf of the Gulf of Lions (GoL), in the Provençal subbasin, 'MEDOC'-area, and in the Ligurian subbasin (e.g., Voorhis and Webb, 1970; Gascard, 1973; Schott and Leaman, 1991). Schroeder et al. (2008) and Smith et al. (2008) suggested a shift of larger convection from the former to the latter subbasin in 2006. This process is typified by O($10^2$-$10^3$ m) horizontal radius for downward motion "plumes" and 10-100 times larger upward motion areas (Marshall and Schott, 1999).

ii) Advection due to horizontal density gradient, frontal zones, such as confirmed in the upper 1000 m off Nice (Stemmann et al., 2008), and mesoscale eddies, such as occur in the Algerian subbasin (van Haren et al., 2006). This process is typified by strong downward currents in an O($10^3$ m) wide rim around eddies on the perimeter, O($10^5$ m) radius, and upward motions in the rest of the area (van Haren et al., 2006).

Here, we report on Acoustic Doppler Current Profiler ADCP-data (RDI, 1992) measuring temperature, echo intensity and current in all three Cartesian components: East-West (u), North-South (v), vertical (w), focussing on episodic relatively large downward w. The echo intensity data are interpreted qualitatively as suspended material, mainly zooplankton. These data are compared with certain observations from PMTs, which are related to bioluminescence. The combination of these data sets is unique and the aim is to better understand the processes governing variability in the deep-sea plankton abundance.

## 2. Materials and methods

In spring 2005, the ANTARES Collaboration deployed and operated a so-called Mini Instrumentation Line equipped with Optical Modules (MILOM) at the site 42°48′N, 06°10′E, 2475 m water depth (Aguilar et al., 2006) (Fig. 1). In March 2006 the first detector line became operational (Ageron et al., 2009). It was placed 78 m from MILOM. Optical data from both lines are used here. The MILOM consisted of an instrumented releasable anchor and of three storeys located at 100, 117 and 169 m above the sea bed. It was equipped with four OMs: a triplet of OMs on the middle storey and a single OM on the upper storey. The line is thus quite like a typical oceanographic mooring, except for the power and data connection to shore. The shoreward data-transport was frequently interrupted in the first half of 2006, but was fine afterwards (Fig. 2).

A downward-looking 300 kHz, four-beam Teledyne RDI-ADCP was mounted on the upper storey of MILOM. The ADCP sampled data-ensembles in 50 vertical bins of 2.5 m every 10 minutes. In this apparatus the beam slant is 20° to the vertical. This leads to current estimates that are averages over horizontal beam spreads of 3-80 m as a function of the vertical range. It also implies that horizontal currents are approximately four times less accurately determined than vertical currents. This is not a problem for determining large-scale ocean currents, which have an aspect ratio of about 1:1000. As a result, attention is given below on the adequate estimating of vertical currents, even though the aspect ratio becomes 1:10 or larger for internal wave motions.

As the ADCP operates a $4^{th}$ beam that is redundant for 3-axis current measurements, it offers an extra 'error' velocity (e) that is composed of the difference between two w-values estimated from the independent beam pairs (RDI, 1992). Due to this definition e is valid as an error estimate for w in all coordinate frames. Thus, reasonable estimates are obtained for errors in w that include horizontal current inhomogeneities over the beam spread (van Haren et al., 1994).

ADCP data are corrected internally by the instrument, using tilt (θ) and heading attitude sensors. Such correction is needed every time-step as the instrument only measures Doppler-shifts ("currents") in the direction of its beams, and the decomposition into [u, v, w] needs to be computed correctly, also for moving instruments. If the attitude sensors are not properly calibrated they can cause bias, especially of [u, v] into w. The observed tilt (Fig. 2), more specifically tan(θ), shows a significant linear relationship with the current amplitude squared $|U|^2=u^2+v^2$ (not shown). This confirms a mooring under current drag (e.g., van Haren, 1996). As will be demonstrated in Section 3, although w and -|U| occasionally have similar temporal variations, differences do occur so that the overall coherence is not significant (not shown).

As the tilt sensors have an accuracy of 0.5°, substantial bias in w is unlikely to happen. A number of independent checks have been performed to further verify this condition.

1) The 0.5° tilt accuracy may in worst case cause a bias of $-3\times10^{-3}$ m $s^{-1}$. Such w-value has been observed using upward looking 75 kHz ADCPs (Schott et al., 1991; van Haren et al., 2006). Part of this w can be attributed to sinking materials, although their typical downward speeds are commonly $-1\times10^{-3}$ m $s^{-1}$ or less (Passow, 1991; Lampitt et al., 1993). In the present 300 kHz downward looking ADCP data a mean of w = 0 is found when currents are weak, e.g. days 146-147, 253-256, 321-323 (see Section 3).

2) The observed total tilt is 2-3° and nearly constant with time (Fig. 2a). Supposing this value is completely biased, and accounting for the slow rotation of the ADCP (Fig. 2b), it is possible to compute an artificial w″ (Fig. 2c; red curve) as influence of horizontal current components [u, v]. It is clearly visible that this w″ has smaller amplitude than the measured w, much smaller standard deviation and frequently has an opposite sign.

3) A more robust verification comes from considering inertial motions which are particularly strong in the area and which would result in artificial w″ being in-phase with u or v. However, as observed in Fig. 2c the much smaller standard deviation comprising internal waves including inertial motions is poorly represented in w″ and the coherence between w and u, v or |U| at this and other frequencies was not statistically significant (not shown). This implies no relevant correspondence between horizontal and vertical motions. As an example, relatively weak horizontal motions other than inertial can be considered (Fig. 3a). When u, v, |U| are near-zero, large negative w may occur (e.g., day 361.9) as well as near-zero w (e.g., day 365.5). This happens occasionally at the larger time scale of a day or more, as well as at the inertial and shorter time scales. |U| and w are seen in-phase as well as out-of-phase in this short record, which cannot be attributed to tilt-bias as then they should always be in-phase. It is noted that vertical component inertial motions are not negligible in near-homogeneous waters and have been observed before in the Provençal subbasin further to the south (van Haren and Millot, 2005).

4) Vertical inertial motions seem mostly uniform over the vertical ADCP-range, as has been observed previously (van Haren and Millot, 2005) in near-homogeneous waters where the buoyancy frequency N is of the same order of magnitude as the inertial frequency f. Such weak stratification is commonly found in CTD-profiles from the basin, but small variations in the vertical do occur (Schroeder et al., 2008; van Haren and Millot, 2009). For instance, when near-bottom stratification (N equaling a few times f) passes the mooring, associated vertical motions have negative maxima well away from the ADCP (Fig. 3b). This also cannot be explained from bias errors in tilt-sensor data, which would affect w equally at all depth levels.

As ADCPs rely completely on the reflection of sound on 'particles' in the water, larger than about 0.003 m at 300 kHz (RDI, 1992), they sample variations in these reflections as 'echo intensity' (I). Part of the I-variation with depth is the inevitable acoustic energy loss in water due to beam spread and chemical reactions (RDI, 1992). A simple method to correct for sound loss is the computation of a 'relative echo intensity', $dI=I-I_{min}$, by subtracting the minimum $I_{min}$ over the entire period of time from the original signal at each depth. When the acoustic signal drops below noise level due to low scatter amounts, dI = 0. A single-frequency instrument cannot be used to distinguish the cause of variations in dI with time. The origin of variations ranges from changes in shape and species to number of particles. Most often however, variations in dI imply variations in the number of particles passing through the beams. A 300-kHz ADCP is sensitive to particles like large suspended flocs of material and especially zooplankton that have sizes $>10^{-3}$ m, or larger animals. It is not sensitive to bacteria and phytoplankton, which have typical sizes $O(10^{-5}-10^{-4}$ m) or less.

## 3. Results

In March 2006, the PMTs counting rates, which at low levels are mainly due to $^{40}$K-decay and to bioluminescent bacteria, suddenly increased by a factor of 10 or more (Fig. 4a). Low levels are observed before day 67 and after day 170, except for moderately intense short episodes. Sudden large increases are apparent for 30 days after day 67 and high levels are maintained until about day 170. Similar observations were made using different PMTs on MILOM and line 1 (Ageron et al., 2009). An increase in counting rate is usually attributed to higher levels of bioluminescence. It is noted that bacterial light is a steady glow and that faunal bioluminescence is characterized by non steady state flashes, e.g., when animals collide with a structure (Priede et al., 2008; Haddock et al., 2010). A larger current magnitude provides more energetic collisions and expected higher rates of bioluminescence, as has been observed in an average sense at ANTARES (Brunner, 2011).

### *3.1. Yearlong overview of strong variations in relative echo intensity, optics and currents*

The ADCP's dI suddenly increased by a factor of about 10 on day 68 (Fig. 4b). Relatively large acoustic reflections occurred until day 190, and episodically later in the year. Larger dI coincide generally with increased PMT counting rates. However, it is noted that optical and acoustic data may imply variations of different origin. The former are sensitive to variations due to distributed light sources, such as luminescent bacteria or zooplankton species. The latter is mostly sensitive to echos due to accumulation of zooplankton and higher order species, not necessarily light-emitting objects. Associated with increases in dI are: large downward w (Fig. 4d) and generally increased horizontal currents, although not necessarily always, e.g., days 85-100 (Fig. 4c). Aside from the period between days 70 and 140 of strongly enhanced dI, large negative w periods of typically 10-30 days of similar but somewhat weaker absolute values are observed later in the year as well, e.g., days 220-230 and 285-300. Variations with time may be more clearly seen in series from a particular depth, e.g., 2320 m (Fig. 5). Large negative w and larger dI and |U| are occasionally accompanied by increases in temperature, but the correlation is ambiguous despite the tendency of large |U| with small T (Fig. 5a). Coherence is not statistically significant between w and |U|, implying w to have partially independent sources. The measurement of e shows that it has a mean of about zero and standard deviation of noise of 0.002 m $s^{-1}$, commensurate with the manufacturer's error estimate in w for given ensemble averaging (Fig. 5c). As a result, current inhomogeneities over the beam spread are not causing any negative bias in w and its apparent noise is mostly due to high-frequency internal waves and small-scale convection.

### *3.2. Details of large downward motion and associated variations*

First focusing on winter-spring of 2006, it is seen that variations in negative w and increased optical measurements, dI and |U| also occur at shorter periods of 1-10 days (Fig. 5). From day 70 to 80, the mean downward motion was about 0.01 m $s^{-1}$ and occasionally exceeded 0.025 m $s^{-1}$ (Fig. 2), i.e. 2000 m $day^{-1}$. The w are best followed by -dI, whilst the optics and |U| show a slightly better correspondence. As for the lack of statistically significant coherence found here, it is noted that acoustic reflections and optical measurements are non-conservative properties, which not only depend on (advective) currents but also on particle sizes and biological influences. The w contains a lot of high-frequency variations that are not noise, but internal waves near the buoyancy period of a few hours. It is noted that these values of w are measured relatively close to the sea bottom, although still well above the frictional bottom boundary layer.

Such large downward motions cannot be associated with sinking particles like heavy diatoms and faecal pellets, whose speeds are 1-2 orders of magnitude smaller (Passow, 1991; Lampitt et al., 1993). They also cannot be associated with zooplankton migration. Zooplankton moves at such speeds, but down and up, in various cycles including a diurnal and a seasonal cycle, in the latter going up in spring (van Haren, 2007).

### *3.3. Details of variations at time-scales of meanders, eddies, jets*

Similar, although less intense, w-variations are observed later in the record (Fig. 6). The optical measurements episodically exceed 300 kHz, roughly with the same 20±10 days periodicity as enhanced w, dI and |U|. Due to warming from May onwards the stratification prevents any deep convection, so that this process definitely cannot explain episodic large summer-autumnal downward currents. Similarly, cascading events in nearby canyons transporting debris down are not expected to occur at a 10-20 days periodicity. Recent sediment and [horizontal] current research (Khripounoff et al., 2009) showed that in the Var-canyon, the nearest canyon upstream of ANTARES, a distinction occurred between resuspension higher-up in the canyon and near its foot. The deep resuspension was mainly attributed to the "regularly" meandering NC influence rather than to flash floods. Locally, in the weakly stratified deep layers [slanted] convection may persist throughout the year associated with sub-mesoscale eddies (Testor and Gascard, 2006) and/or near-inertial internal waves (van Haren and Millot, 2009). However, such processes are not known for 20-day quasi-periodicity.

Progressive Vector Diagrams (PVDs) constructed via time integration of horizontal particle velocities using the deep ADCP data show predominant westward 'displacement' between days 69 and 79, preceded by northward and followed by southward displacements (Fig. 7a). Although ambiguous, this could be interpreted as due to the passage of a mesoscale meander or clockwise eddy passing with its core between the ANTARES site and the coast during westward propagation with the prevailing NC, such as observed previously (Crépon et al., 1982).

Generally, the baroclinic unstable, meandering NC passes inshore of the ANTARES site as is observed from sea surface satellite images (e.g., Fig. 7b). Particularly on day 68 we observe a strong baroclinic instability forming a vortex pair or dipole just to the East of the ANTARES site (Fig. 7b), with a seaward central jet. This is visualized in the large change in surface chlorophyll ('colour'). The dimensions are 40x80 km, about twice the amplitude and wavelength of typical NC-instabilities that are visible to the West of the dipole and which occur at 10-20 day intervals. The size of the dipole compares well with previous observations affecting surface plankton in the Atlantic Ocean (Gower et al., 1980). Many good satellite images could not be obtained over the following days as a result of cloudiness, however the boundary in surface chlorophyll seemed more or less stationary over the ANTARES site for about 10 days. While the dipole clearly developed from an instability of the NC, images show intense mesoscale eddy activity in the central basin that occasionally come close to the NC at the ANTARES site (Fig. 7c).

We speculate that one can relate sea surface satellite images with deep-sea ADCP observations, as seems justifiable for quasi-barotropic Gulf Stream and Algerian mesoscale eddies, with >2000 m vertical extent (Richardson, 1983; Millot et al., 1997; van Haren et al., 2006), and also for sub-mesoscale eddies with >1000 m vertical uniformity (Testor and Gascard, 2006). For the presumed more baroclinic NC, the present observations compare to some extent with models on dipoles, which have near-zero phase speed due to their interaction with the sheared current deep below (Griffiths and Pearce, 1985; Crépon et al., 1989). As a result, a particular area may receive persistent vertical flux of material during the lifetime of a jet or mesoscale meander or eddy (we cannot distinguish between them from an ADCP-record). Later in the ADCP-record we do find indication for more meanders or eddies, associated with episodes of relatively large downward vertical currents. The PVDs show eddies of a clockwise nature, but it is noted that eddy interpretation from PVD can be ambiguous when no other information is available.

## 4. Discussion

The beginning of 2006 was characterized by particularly strong convection and deep dense water formation observed in the Provençal MEDOC-area, in particular in the Ligurian-subbasin during two major periods, January and March/April, and which lasted well into spring (Schroeder et al., 2008; Smith et al., 2008). However, it is unlikely that the present large current, acoustic and optical variations, observed over the course of a year, are exclusively associated with dense water formation processes. In the first instance, temperature increases are considered evidence of dense water formation in the Ligurian subbasin (Schroeder et al., 2008; Smith et al., 2008). However in this study downward motions show ambiguous correspondence with temperature increases. Secondly, episodic downward motions were observed throughout the year, not only in winter and early spring.

In contrast, the NC is a permanent, non-random current, as is its meandering activity, although modulated by seasonal variation and (re)inforced by dense water formation (Crépon et al., 1989). The meandering NC-rim is a good candidate to cause temporal variations in current, acoustic reflection and optical measurements at the ANTARES site periodically all year long, when forced to great depths.

As a result, it is hypothesized that active mesoscale motions could cause the observed large variations in acoustic reflection, optical measurements, horizontal and downward motions. As such motions are generated via instabilities of the coastal-continental topographic NC-system in response to atmospheric forcing above the deep open subbasin, they follow periods of exceptional dense water formation and last for months.

The present acoustical observations support the optical observations of the ANTARES array that large amounts of particles are transported downward from higher up, commonly with enlarged horizontal motions thus spiraling advectively down, resulting in episodically high counting rates in the PMTs. As both acoustics and optical sensors respond more or less during the same episodes and because acoustics are insensitive to bacteria, an important contribution to bioluminescence can be ascribed to zooplankton, or, perhaps though unlikely, to large suspended material like faecal pellets (Andrews et al., 1984) carrying luminescent bacteria. These observations provide an indication that mesoscale meanders or eddies, generated by unstable currents, may create episodic increases in the flux of fresh organic material to the deep sea.

*Acknowledgments.* Satellite images are obtained from NASA/MODIS, processed by Ifremer and available on the Nausicaa/MarCoast server. We thank Michel Crépon, Claude Millot and anonymous referees for commenting an earlier draft. The authors acknowledge the financial support of the funding agencies: Centre National de la Recherche Scientifique (CNRS), Commissariat á l'Energie Atomique (CEA), Commission Européenne (FEDER fund and Marie Curie Program), Région Alsace (contrat CPER), Région Provence-Alpes-Côte d'Azur, Département du Var and Ville de La Seyne-sur-Mer, in France; Bundesministerium für Bildung und Forschung (BMBF), in Germany; Istituto Nazionale di

Fisica Nucleare (INFN), in Italy; Stichting voor Fundamenteel Onderzoek der Materie (FOM), Nederlandse organisatie voor Wetenschappelijk Onderzoek (NWO), in the Netherlands; Federal Agency for Science and Innovation (Rosnauka), in Russia; National Authority for Scientific Research (ANCS) in Romania; Ministerio de Ciencia e Innovación (MICINN), in Spain. We also acknowledge the technical support of Ifremer, AIM and Foselev Marine for the sea operation and the CC-IN2P3 for the computing facilities.

**Figure 1**. ANTARES site (red star) roughly located on the northern edge of the border between the Ligurian (L) and Provençal (P) subbasins, western basin Mediterranean Sea. Indications are given of the rim of the Northern Current NC (solid line) and areas of dense-water formation (dw). Isobaths every 500 m between [-500, -2500]m.

**Figure 2**. ADCP's tilt, heading and vertical current comparison. The time convention is January 1, 12.00 UTC = day 0.5, 2006. a. Pitch (blue), roll (black) and total tilt (green). b. Compass-heading. c. Raw vertical current observed at 15 m below ADCP (blue), in comparison with artificial w″ computed from horizontal currents at the same depth using tilt-sensor data (red).

**Figure 3**. a. Detail comparison between horizontal current components (blue: u; black: v; green: -|U|) and vertical current (red: scale to right) measured at 30 m below the ADCP in winter 2006/2007. b. Depth-time series of vertical motions, which show largest amplitudes (most negative values) at about 70 m below the ADCP, in this example. Days in 2007 are +365.

**Figure 4**. a. Optical counting rate observed 50 m below the ADCP at MILOM (blue) and on Line 1 (red) as a function of time. b.-d. Raw MILOM-ADCP data, time-depth series. In all panels the vertical white lines indicate absence of data. The two horizontal lines at 2350 and 2365 m are direct sound reflections from two storeys below the ADCP. b. Relative echo amplitude from a beam, limited to [0, 12] dB. c. Current amplitude, between [0, 0.2] m $s^{-1}$. d. Vertical current, between [-0.01, 0.01] m $s^{-1}$. In c., d. useful data are available down to about 2390 m, and to about 2420 m between days 70 and 145 when echos are large. At depths further from the ADCP than this the signal to noise ratio approached one due to radial sound loss and lack of scatterers. This results in acoustic levels at minimum constant noise level $I_{min}(z)$ and hence dI = 0 (dark-blue).

**Figure 5**. Late winter-spring 2006 time series. Applied smoothing is using a 20-points running mean. a. Horizontal current amplitude at 2320 m (black; smoothed data) and temperature (green; smoothed) measured at the ADCP. b. Relative sound-echo amplitude (black; smoothed) and PMT baseline data from 2350 m (red; raw data). c. Vertical current (black: smoothed) and error velocity (green: smoothed).

**Figure 6**. As Fig. 5, but for summer-autumn 2006 (early winter 2007) time series.

**Figure 7**. a. Progressive Vector Diagram of integrated Eulerian horizontal currents observed at 2320 m. In black the total 2006-time series that start at (0,0), in colours portions between the days indicated. b. Satellite image of false-coloured near-surface chlorophyll-a on day 68. The strongly meandering rim of NC is approximately delineated by the sharp change from light- dark-blue near the coast. c. As b., but for near-surface suspended particulate matter on day 87. Here, the darker-blue colours are near the coast, but as before the strong colour contrast delineates the meandering NC-rim. In b. and c. the ANTARES site is marked by a cross. Note the images are used qualitatively for pattern recognition.

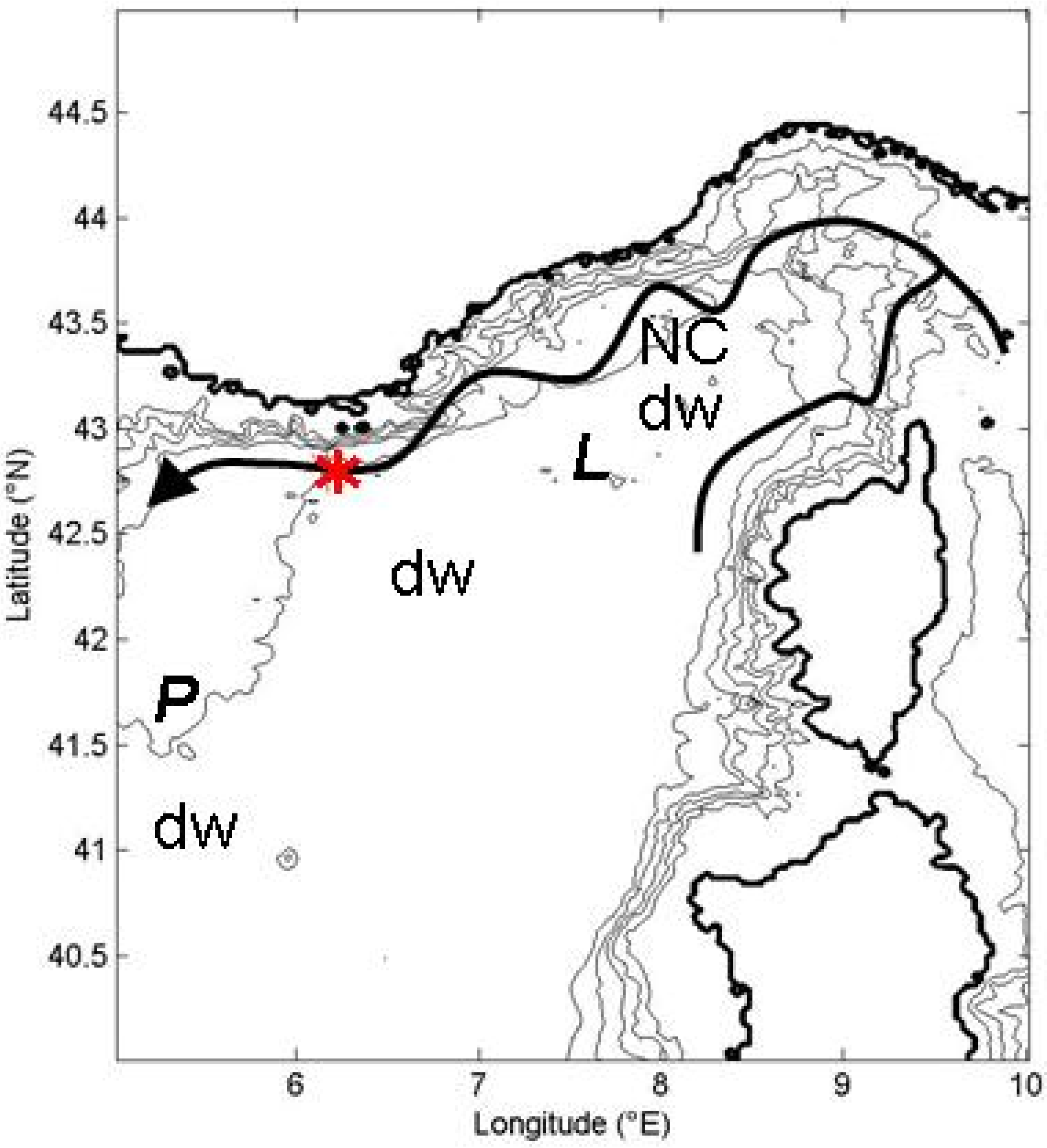

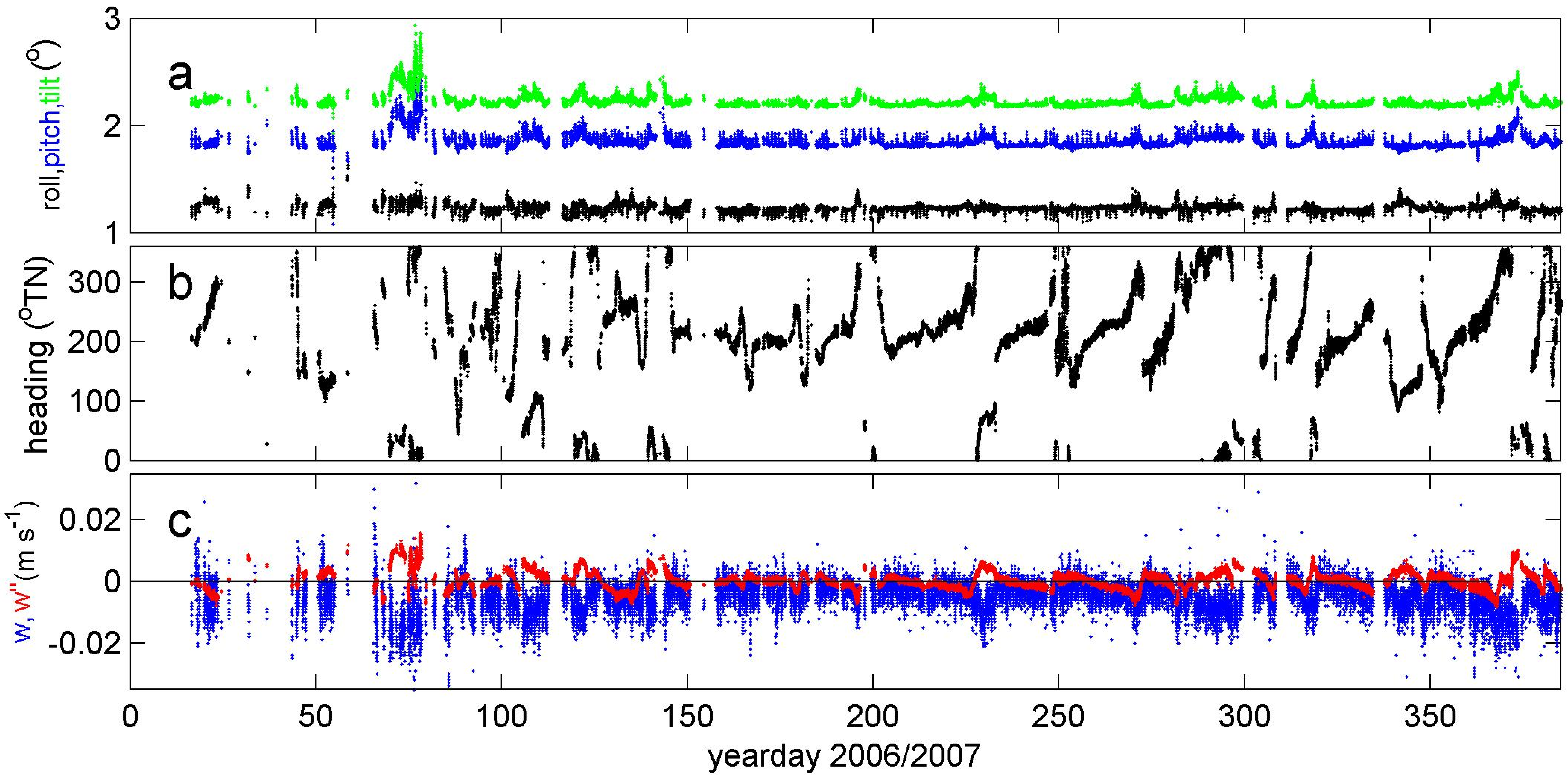

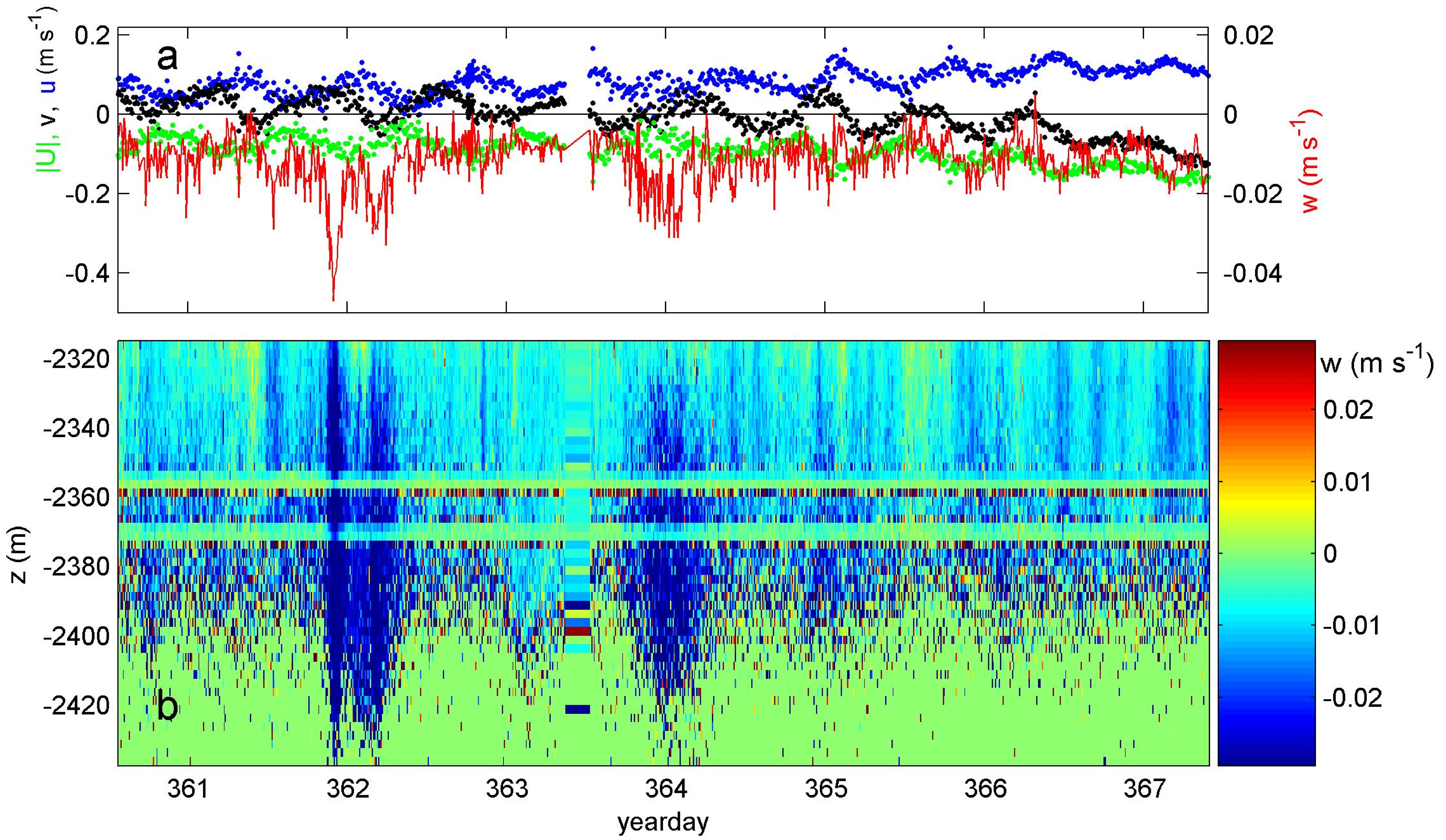

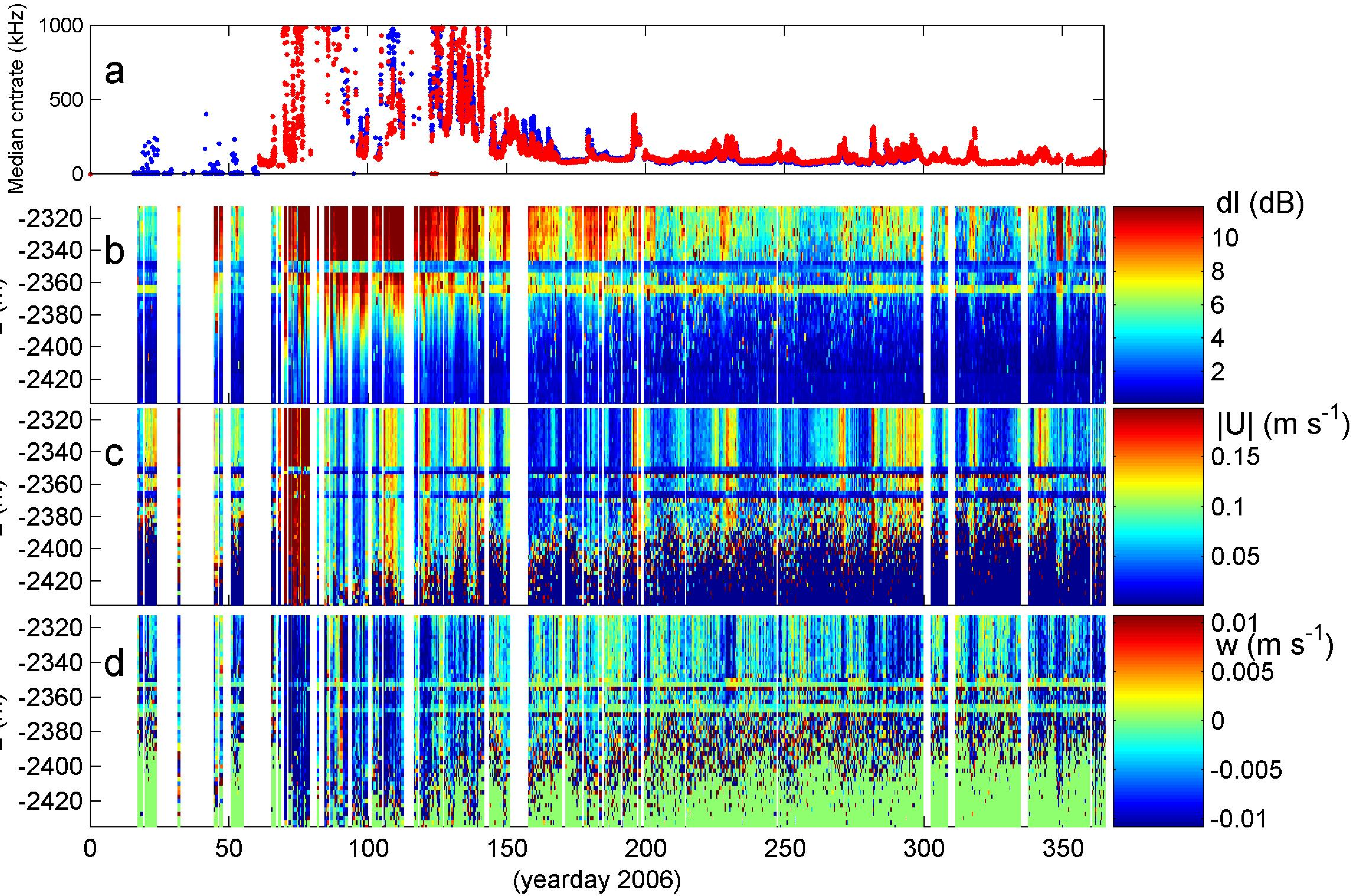

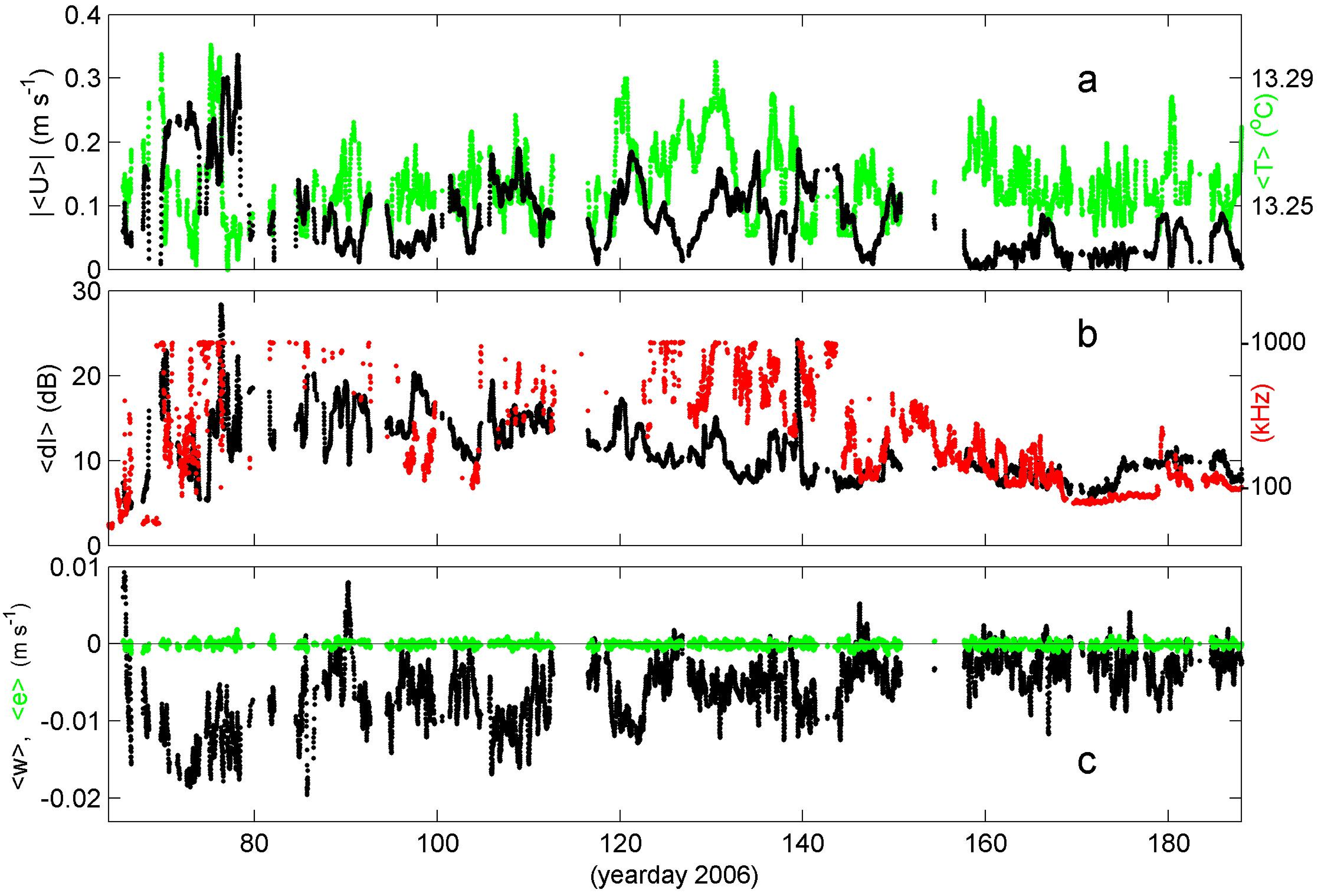

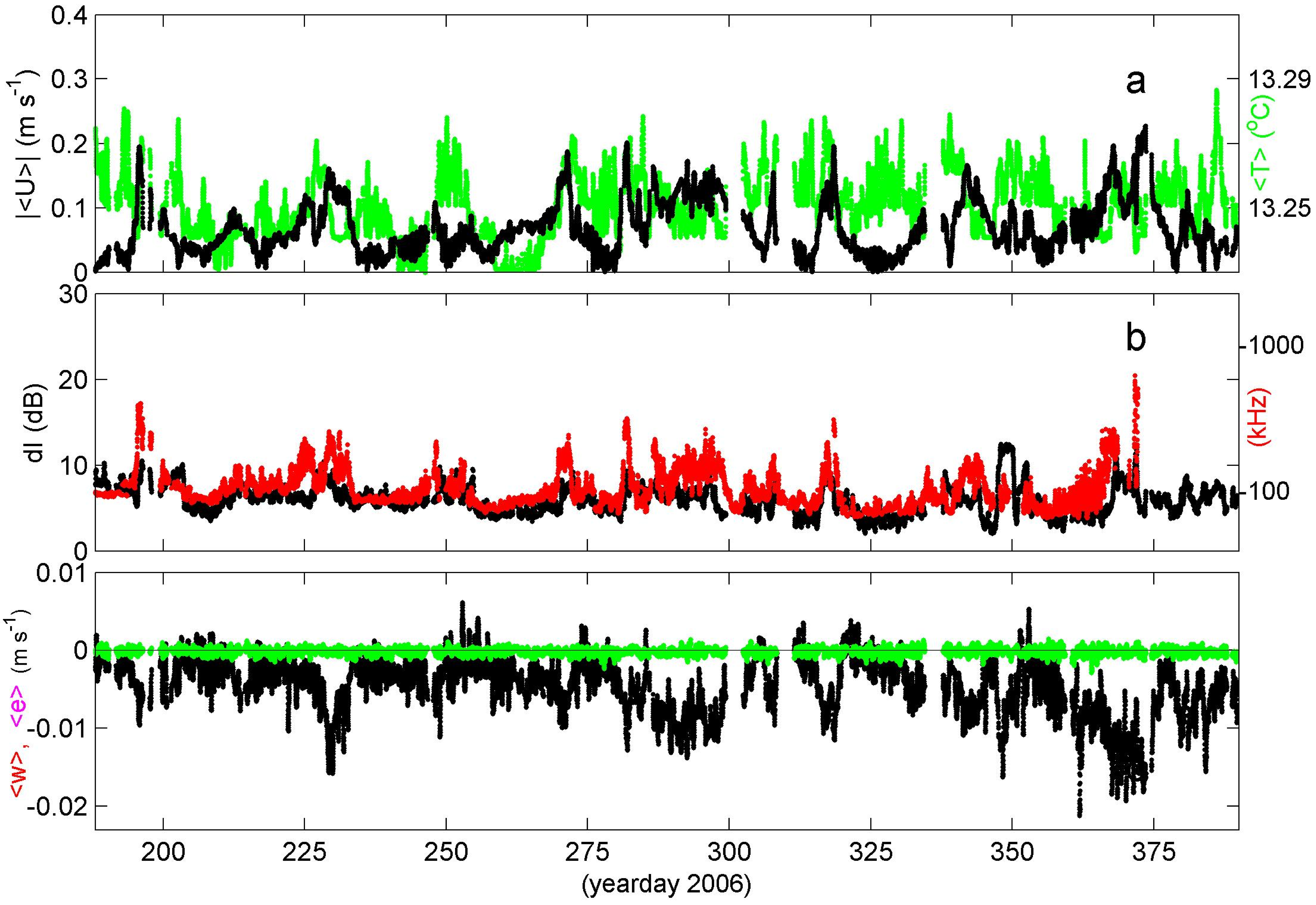

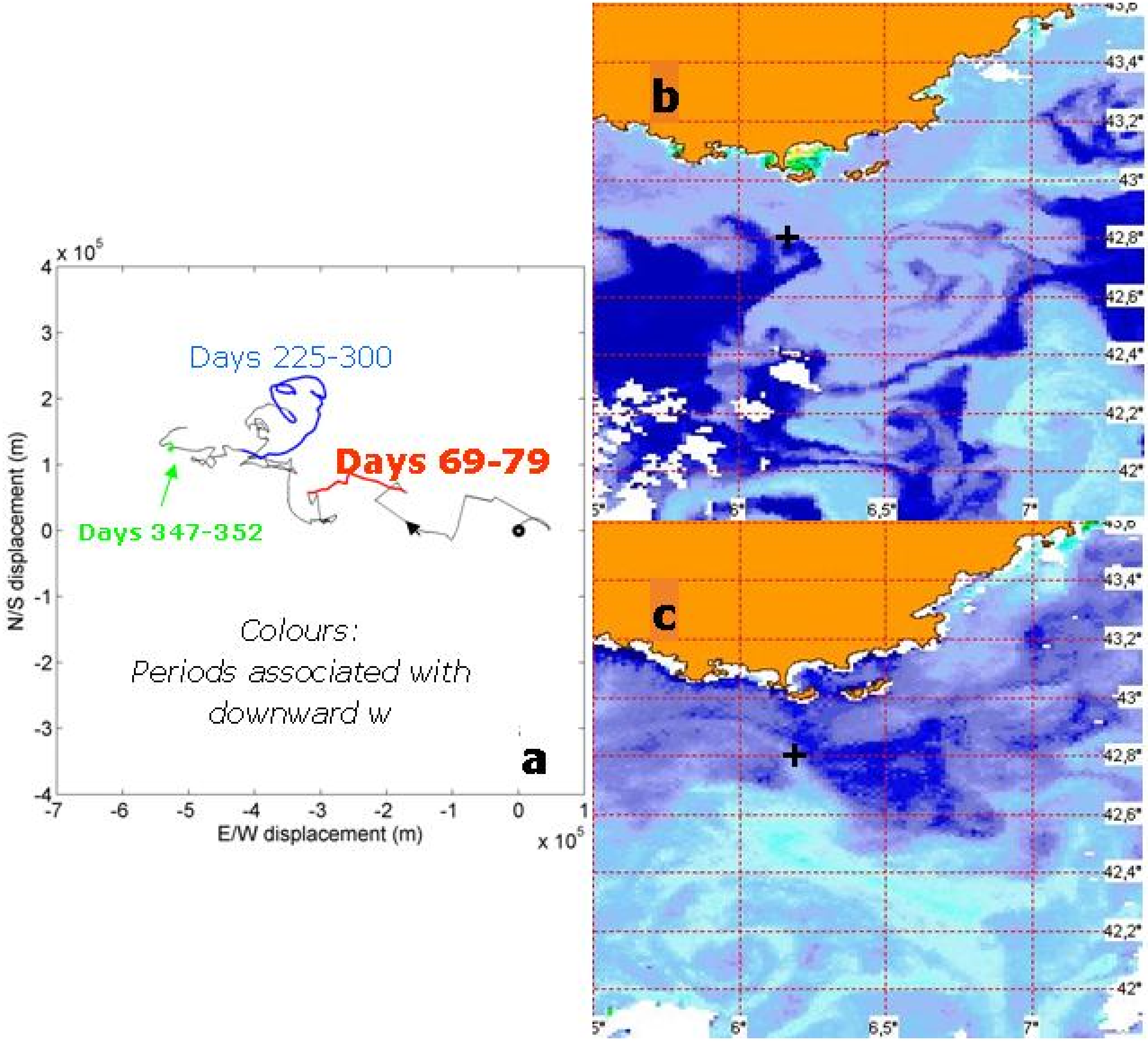